# Nonlinear dielectric response of Debye, $\alpha$, and $\beta$ relaxation in 1-propanol

Th. Bauer, M. Michl, P. Lunkenheimer[*], A. Loidl

*Experimental Physics V, Center for Electronic Correlations and Magnetism, University of Augsburg, 86135 Augsburg, Germany*

ABSTRACT

We present nonlinear dielectric measurements of glass-forming 1-propanol, a prototypical example for the monohydroxy alcohols that are known to exhibit unusual relaxation dynamics, namely an additional Debye relaxation, slower than the structural $\alpha$ relaxation. Applying high ac fields of 468 kV/cm allows for a detailed investigation of the nonlinear properties of all three relaxation processes occurring in 1-propanol, namely the Debye, $\alpha$, and $\beta$ relaxation. Both the field-induced variations of dielectric constant and loss are reported. Polarization saturation and the absorption of field energy govern the findings in the Debye-relaxation regime, well consistent with the suggested cluster-like nature of the relaxing entities. The behavior of the $\alpha$ relaxation is in good accord with the expectations for a heterogeneous relaxation scenario. Finally, the Johari-Goldstein $\beta$-relaxation in 1-propanol seems to exhibit no or only weak field dependence, in agreement with recent findings for the excess wing of canonical glass formers.

**1. Introduction**

In recent years, the investigation of the nonlinear dielectric properties of glass-forming materials has gained increasing interest (see, e.g., [1,2,3,4,5,6,7,8,9,10,11,12,13]). In conventional dielectric spectroscopy, the linear response of a material to moderate electrical fields is detected [14,15,16]. In contrast, the application of high fields up to several 100 kV/cm drives the investigated glass former into the nonlinear regime and can reveal important additional information about the glass transition and the glassy state of matter. For example, dielectric hole-burning experiments have first proven the heterogeneous nature of glassy dynamics [17]. Further valuable information on the dynamic heterogeneity of glass formers was gathered by detecting the alteration of the permittivity under high ac fields [1,3]. Moreover, based on a model by Bouchaud, Biroli, and coworkers [18,19], the increase of molecular cooperativity when approaching the glass transition was investigated by measurements of the higher-order susceptibility $\chi_3$ [7,8,9,13].

Compared to most of the dipolar glass-forming liquids that are typically investigated by dielectric spectroscopy, many monohydroxy-alcohols were found to show unusual relaxation dynamics: The slowest and, in most cases, dominating relaxation process revealed in their low-field dielectric spectra does not correspond to the structural relaxation process, i.e. the molecular motion governing, e.g., viscous flow [20,21,22,23]. Instead, this so-called Debye process is usually ascribed to the much slower motions of clusters (chain or ring-like structures) formed by several hydrogen-bonded alcohol molecules [10,23,24]. However, the details of these relaxational motions still need to be clarified. In spectra of the dielectric loss, this process shows up as a peak whose spectral shape can be well described by the Debye function, $\varepsilon''(\nu) = \Delta\varepsilon \, \omega \tau_D / [1 + (\omega\tau_D)^2]$, where $\omega = 2\pi\nu$ is the angular frequency, $\Delta\varepsilon$ is the relaxation strength, and $\tau_D$ the relaxation time. In contrast, the loss peak arising from the structural $\alpha$ relaxation, which is the dominating spectral feature in most other glass-forming liquids and which also contributes to $\varepsilon''(\nu)$ at $\omega > 1/\tau_D$ in monohydroxy alcohols, usually does not follow this function. Instead it is significantly

---

[*] Corresponding author. *E-mail address:* Peter.Lunkenheimer@Physik.Uni-Augsburg.de (P. Lunkenheimer)

broadened and often asymmetrically shaped. This can be ascribed to the heterogeneous nature of glassy dynamics [25], leading to a distribution of relaxation times, i.e. each single molecule relaxes in accord with the Debye theory, but the relaxation times are different for different molecules. However, the molecular dynamics in the environment of the mentioned supramolecular clusters in the monohydroxy alcohols, which is dominated by the $\alpha$-relaxation time $\tau_\alpha \ll \tau_D$, is much faster than the cluster motion itself. Thus, any heterogeneity in the material is blurred by these faster molecular fluctuations and a monodispersive Debye-shaped loss peak is observed.

In the present work, we investigate the nonlinear dielectric response of 1-propanol, a prototypical material that was among the first examples, where the non-canonical behavior of monohydroxy alcohols was unequivocally demonstrated [20]. In earlier works by R. Richert and coworkers, strong variations in the nonlinear properties of different monohydroxy alcohols were found [6,10,11]. In [11], this finding was ascribed to differences in the ability of high electrical fields to affect the equilibrium of cluster shapes fluctuating between polar open-chain and nonpolar ring-like structures. However, 1-propanol seems to be unaffected by this mechanism [11] and, thus, is an ideal candidate to investigate in detail the nonlinear behavior for the single-dispersive case, lacking any heterogeneity. Nonlinear dielectric experiments on canonical glass formers as glycerol, checking for the field-induced variation of the permittivity, can be well understood considering dynamical heterogeneity [1,3]. Moreover, nonlinear glassy dynamics can also be interpreted in terms of cooperativity effects [7,8,13,18], another important aspect often invoked to explain the peculiarities of the supercooled and glassy state of matter. 1-propanol represents a much simpler system where the first mentioned hallmark feature of glassy dynamics (heterogeneity) seems to be absent and the second (cooperativity) should play a smaller role only, as cluster-cluster interactions can be expected to be rarer than the intermolecular interactions in canonical glass formers. Moreover, extending the investigated frequency range to the region of the $\alpha$ and $\beta$ relaxations will also provide information on the nonlinear behavior of these processes.

In the present work, we report the modification of the dielectric permittivity (real and imaginary part) by the application of high ac fields. The use of microspheres as capacitor-spacer material (see section 2) allows for the application of very high fields of 468 kV/cm. Thus, the obtained permittivity results are of unprecedented precision and cover a broader frequency range than most earlier nonlinear investigations of glass forming materials.

## 2. Experimental procedures

The measurements were performed using a frequency-response analyzer in combination with a high-voltage booster "HVB 300", both from Novocontrol Technologies, enabling measurements with peak voltages up to 150 V at frequencies up to about 100 kHz. The sample material (1-propanol of 99.7 % purity, anhydrous) was purchased from Aldrich and mixed with 0.05 % silica microspheres (2.87 μm average diameter, monodisperse, plain) from Corpuscular Inc. When putting the sample between two lapped and highly polished stainless steel plates, these dielectrically neutral microspheres act as spacing material, leading to an extremely small plate distance enabling the application of very high fields of up to 468 kV/cm. A sample thickness of 3.2 μm was deduced from a comparison of the absolute values of $\varepsilon''$ with the published low-field results from [20,21]. For a verification of the obtained results, additional measurements with glass-fiber spacers of 30 μm diameter were carried out, using a high-voltage booster "HVB 4000", reaching voltages up to 2000 V and frequencies up to about 1 kHz. Similar to the procedure reported in Refs. [1,3], at each frequency we performed successive high- and low-field measurements, separated by a waiting time. To minimize effects from phonon heating, few high-field oscillations (150 V, 468 kV/cm) were applied, followed by a cooling period achieved by applying a series of "waiting-oscillations" with 0.7 V only. Subsequently, a low-field measurement with 4.5 V (14 kV/cm) was carried out. At low frequencies typically 8 high-voltage cycles were applied while at higher frequencies the cycle number was larger (and determined by the processing speed of the experimental setup). For example, at 100 Hz, 30-50 cycles were applied, corresponding to a measurement time of 0.3 - 0.5 s and for $\nu \geq 1$ kHz, the field was always applied for one second. The number of applied "waiting-oscillations" was 27 times higher than the cycle number of the high-field measurement to ensure that the low-field data are not affected by the preceding high-field measurement. For cooling, a closed-cycle refrigeration system (CTI-Cryogenics) was used.

## 3. Results and discussion

### 3.1. Debye relaxation

Fig. 1 shows broadband loss spectra of 1-propanol obtained by conventional low-field dielectric spectroscopy, measured at various temperatures. As demonstrated in [26], where part of these data were already shown, these spectra can be reasonably fitted by the sum of three peak functions. For the curve at 112 K, these peaks are indicated by the lines. The one located at the lowest frequency is the Debye peak and can be well described by the Debye function. The second relaxation corresponds to the structural $\alpha$ relaxation, for which a Cole-Davidson function [27] was used. Finally, 1-propanol also shows a $\beta$ relaxation as also often found in canonical glass formers [28,29]. In the fits it was accounted for by a Cole-Cole function [30]. This relaxation in 1-propanol is commonly assumed [20,21,31] to be of Johari-Goldstein type [32]. Recently the secondary relaxations found in several other monohydroxy alcohols were also assigned to this class of dynamic processes [11]. Johari-Goldstein relaxations are assumed to be inherent to the glassy state of matter [32] but, until now, no consensus about their microscopic origin has been reached. The temperature-dependent relaxation times of the three detected relaxation processes, determined from the mentioned fits, are consistent with literature data [20,21] and discussed in detail in Ref. [26]. Both $\tau_\alpha(T)$ and $\tau_D(T)$ significantly deviate from Arrhenius behavior.

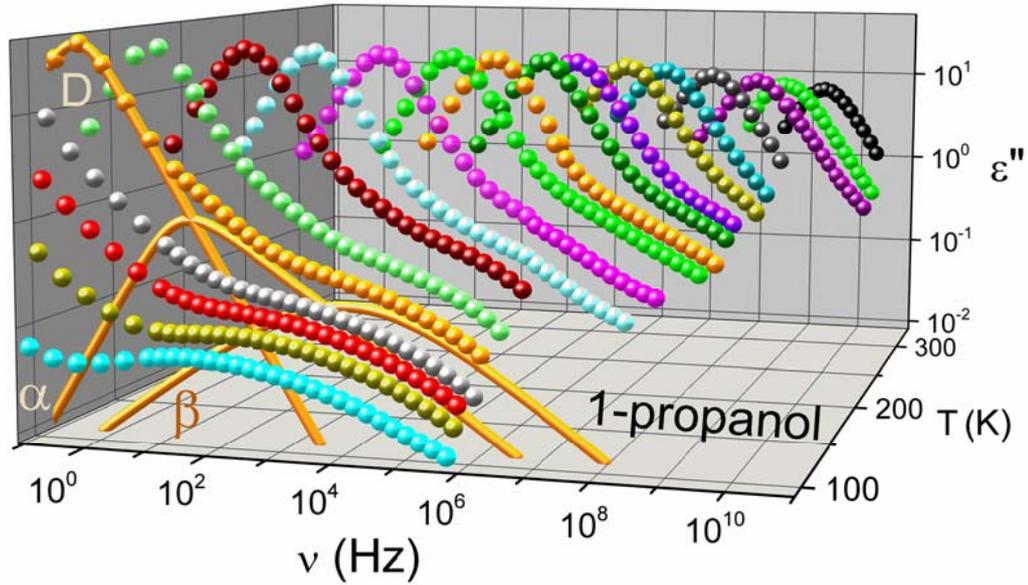

**Fig. 1.** Broadband dielectric loss spectra of glass forming 1-propanol at selected temperatures. The lines demonstrate the composition of the curve at 112 K by three separate relaxations peak arising from Debye (D), $\alpha$, and $\beta$ relaxation as obtained by fits of the data [26].

As an example for the performed measurements of the dielectric loss at low ($E_l$ = 14 kV/cm) and high ac fields ($E_h$ = 468 kV/cm), Fig. 2 shows the spectra at 108 K. The main effect of the application of high ac fields seems to be a shift of the Debye peak to higher frequencies, which leads to a decrease of $\varepsilon''$ at $\nu < \nu_D = 1/(2\pi\tau_D)$, where $\nu_D$ is the peak frequency, and an increase at $\nu > \nu_D$ (see insets of Fig. 2). In the region of the $\alpha$ relaxation, a much smaller variation of $\varepsilon''$ is observed while both curves practically coincide in the $\beta$-relaxation regime. More detailed information on the nonlinear behavior can be obtained by plotting the difference of the high- and low-field spectra. Following earlier work [1,12], in Fig. 3(a) we show the quantity $\Delta \ln \varepsilon'' = \ln \varepsilon''(E_h) - \ln \varepsilon''(E_l)$. Indeed a transition from negative to positive $\Delta \ln \varepsilon''(\nu)$ (corresponding to a decrease or increase of $\varepsilon''$ under high field, respectively) occurs. A comparison with the low-field results of $\varepsilon''(\nu)$, plotted in Fig. 3(b), reveals that this zero-crossing does occur close, but not exactly at the $\alpha$-peak frequency, which is indicated by the solid arrows in Fig. 3(a). The maximally reached absolute values in the negative region of $\Delta \ln \varepsilon''(\nu)$ at low frequencies are significantly larger than the maximum values in the positive region. Moreover, there is an increase of $\Delta \ln \varepsilon''(\nu)$ at the lowest frequencies, leading to a minimum in its negative region. Finally, a shoulder is found in the positive region of $\Delta \ln \varepsilon''(\nu)$, followed by a decrease, approaching values close to zero, if the frequency is further increased. Overall, the observed nonlinear behavior of $\varepsilon''(\nu)$ is rather complex and for its explanation a number of different contributions have to be considered as discussed in the following.

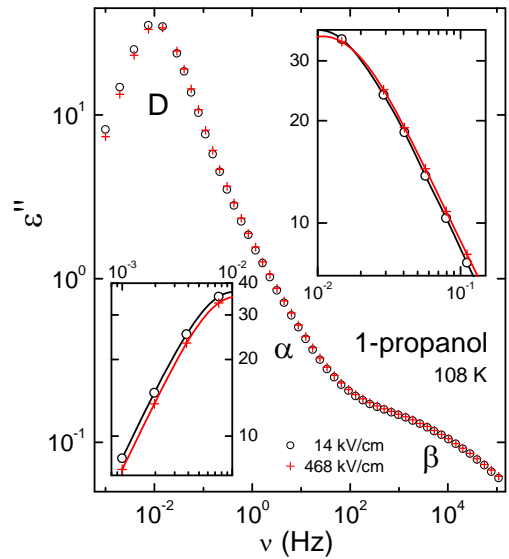

**Fig. 2.** Dielectric loss spectra of 1-propanol measured at 108 K and two different ac fields as indicated in the figure. The insets provide magnified views of the left and right flank of the Debye peak (the lines are guides to the eye).

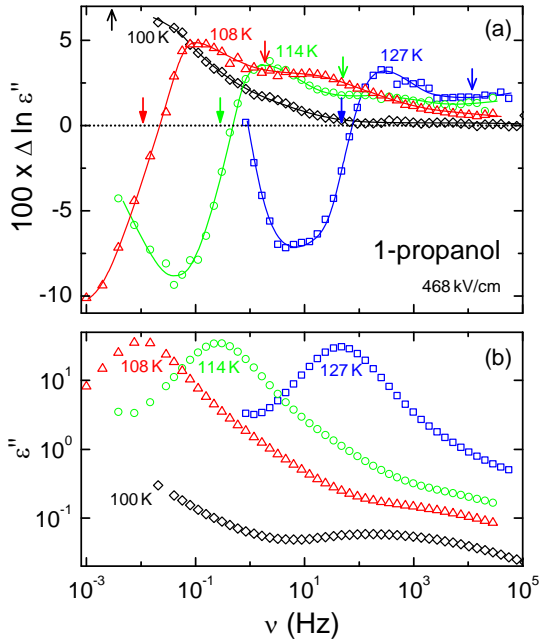

**Fig. 3.** (a) Difference of the logarithms of the loss spectra of 1-propanol, measured with high and low field, plotted for selected temperatures. The solid and open arrows indicate the Debye- und $\alpha$-peak positions, respectively. The lines are shown to guide the eyes. (b) Corresponding low-field loss spectra.

In contrast to the present results on 1-propanol, in conventional dipolar glass formers like glycerol or propylene carbonate a pronounced field-induced variation of $\varepsilon''$ was only found at the high-frequency flank of the main relaxation peak while the loss was only weakly sensitive to high fields below the peak frequency [1,3,12]. This finding can be quantitatively understood within the so-called box model [17,33] if considering the presence of a distribution of relaxation times caused by dynamical heterogeneity [1,3]: Within this scenario it is assumed that the field-induced variation of $\varepsilon''$ is caused by a selective transfer of field energy into the heterogeneous regions. The $\alpha$-relaxation peaks of many glass formers can be fitted by the Cole-Davidson function [15,16,27,29], which is asymmetrically broadened compared to the monodispersive Debye function [27]. The corresponding relaxation-time distribution function is strongly asymmetric and only comprises times $\tau < \tau_\alpha$ [34]. Thus there are no heterogeneous regions with relaxation rates slower than the loss-peak frequency $\nu_p \approx 1/\tau_\alpha$. Therefore only weak absorption will occur for $\nu < \nu_p$ and $\Delta \ln \varepsilon''(\nu)$ is strongly asymmetric with no or only a minor negative contribution at low frequencies [1,3].

As discussed in section 1, for the Debye process of monohydroxy alcohols heterogeneity should play no role. Therefore, naively a simple field-induced shift of the whole Debye peak to higher frequencies due to the heating effect of the field could be expected. In $\Delta \ln \varepsilon''(\nu)$ this would correspond to a transition from a negative to a positive plateau (both of same magnitude) with zero crossing at $\nu = \nu_D$. However, a more detailed analysis revealed that in the Debye case an asymmetric $\Delta \ln \varepsilon''(\nu)$ curve should arise, too, with a much larger amplitude in the positive than in the negative region [6]. Thus, the spectra of $\Delta \ln \varepsilon''(\nu)$ for a monodispersive (homogeneous) and polydispersive (heterogeneous) relaxation are qualitatively similar. Such asymmetric behavior with a small negative and strong positive contribution, following the model prediction, was indeed explicitly demonstrated for the monohydroxy alcohol 2-ethyl-1-butanol [6]. The asymmetry can be made plausible if considering that the absorption of field energy is approximately proportional to the real part of the conductivity $\sigma'$, which is related to the loss via $\sigma' \propto \varepsilon''\nu$. This leads to higher field absorption (and thus heating) at frequencies above $\nu_D$ than below, even for the Debye case as in the monohydroxy alcohols.

Interestingly, $\Delta \ln \varepsilon''(\nu)$ at the Debye peak of 1-propanol (Fig. 3(a)) shows strong negative values at low frequencies, i.e. it behaves significantly different than discussed in the previous paragraph. The positive values found in the region of the high-frequency flank of the Debye peak (cf. Fig. 3(b)) can be ascribed to the mentioned heating effects. In Ref. [6], while no spectra on propanol were shown, a high-frequency amplitude of $\Delta \ln \varepsilon'' = 0.18\%$ was reported for 125 K and 100 kV. If accounting for the quadratic field dependence of the observed nonlinear effects, from our results at 127 K and 1 kHz, we arrive at 0.12 % for this field, which is of similar order of magnitude. However, the strong negative values of $\Delta \ln \varepsilon''(\nu)$ at first glance seem to be at variance with the expectations for a Debye process and they also qualitatively disagree with the results on 2-ethyl-1-butanol [6] mentioned above.

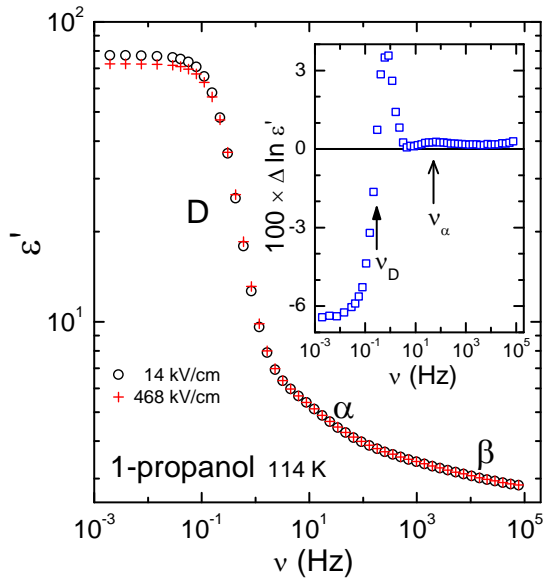

**Fig. 4.** Spectra of the dielectric constant of 1-propanol measured at 108 K and two different ac fields as indicated in the figure. The inset shows the difference of the logarithms of the high- and low-field spectra. The solid and open arrows indicate the Debye- und $\alpha$-peak positions, respectively.

A clue to the origin of this finding is provided by the results on the real part of the permittivity $\varepsilon'$ shown in Fig. 4. As a typical example, the figure presents the $\varepsilon'(\nu)$ results at 114 K as measured at low and high field. The expected steplike decrease of $\varepsilon'(\nu)$, characteristic for dielectric relaxation, is observed. The deviations from a simple symmetric step at high frequencies are due to contributions from the $\alpha$ and $\beta$ relaxations. Obviously, the strongest deviations of the low- and high-field results in Fig. 4 occur in the low-frequency plateau region of $\varepsilon'(\nu)$, which corresponds to the static value $\varepsilon_s$ of the dielectric constant. Such a reduction of $\varepsilon_s$ in high fields is a well-known effect and was treated in detail in many pioneering works on nonlinear dielectric properties (see, e.g., [35,36]). It arises from the saturation of polarization occurring at low frequencies and high fields. As shown, e.g., in [37], this saturation effect (sometimes also termed Langevin effect) also leads to a reduction of the dielectric loss at low frequencies, thus explaining the strong negative values in $\Delta \ln \varepsilon''(\nu)$ at low frequencies, revealed in Fig. 3(a).

In the inset of Fig. 4 the field-induced variation of $\ln \varepsilon'$ is presented, again demonstrating the saturation-induced reduction of the dielectric constant at the lowest frequencies. Just as for $\Delta \ln \varepsilon''(\nu)$ (Fig. 3(a)), the positive peak immediately following the zero crossing of $\Delta \ln \varepsilon'(\nu)$ arises from the absorption of field energy. The overall behavior of the found field-induced variation of $\varepsilon'$ in the Debye-peak region qualitatively agrees with that reported in Ref. [11] for 1-propanol measured at lower fields of 212 kV/cm. In that work, also results on 5-methyl-3-heptanol and 4-methyl-3-heptanol were provided. Qualitatively different behavior of $\varepsilon'(\nu)$ of these monohydroxy alcohols in high fields compared to 1-propanol, namely a positive $\Delta \ln \varepsilon'$ at low frequencies, $\nu < \nu_D$, was found. This was explained by field-induced ring-chain conversions of the molecular clusters at low frequencies [10,11]. For 1-propanol, such effects seem to play no or only a minor role [11]. Indeed, molecular dynamics simulations of 1-propanol resulted in a rather small percentage of molecules participating in ring structures [38].

The magnitude of the reduction of $\varepsilon_s$ caused by the mentioned saturation effect can be calculated, e.g., using the formula [3,39,40]

$$\varepsilon_s(E_h) - \varepsilon_s(E_l) = -\frac{\mu^4}{45\varepsilon_0 V_{mol}(k_B T)^3} \frac{\varepsilon_s^4(\varepsilon_\infty + 2)^4}{(2\varepsilon_s + \varepsilon_\infty)^2(2\varepsilon_s^2 + \varepsilon_\infty^2)}$$

.

Here $\mu$ denotes the dipolar moment, $\varepsilon_0$ the permittivity of free space, $\varepsilon_\infty$ the limiting high-frequency dielectric constant, and $V_{mol}$ the molecular volume. When using $\mu$ and $V_{mol}$ of the 1-propanol molecule, the application of this formula leads to a value that is about a factor of four smaller than the experimentally determined result (Fig. 4). This apparent discrepancy indicates that the Debye process indeed involves molecular clusters with a higher dipolar moment than a single molecule. When assuming that the total dipolar moment and volume of a hydrogen-bound cluster, formed by $n$ molecules, is simply given by $n\mu$ and $nV_{mol}$, respectively, we arrive at an average $n$ of 1.6. This is much smaller than the magnitudes of $n$ varying between 3 and 14 reported for various monohydroxy alcohols [23,24] and essentially would imply dimer formation only. Notably for 1-propanol a rather small average cluster size of $n = 3$ was obtained from Monte Carlo simulations [41]. Moreover, molecular dynamics simulations of a series of linear alcohols from methanol to tridecanol showed that the clusters in propanol contained unusually few molecules [38]. However, this study also suggests that a considerable part of the clusters in propanol is of branched nature. Obviously the above assumptions for the calculation of $n$ are too oversimplified to enable an exact statement about its magnitude. In any case, the present results clearly demonstrate that the Debye process is not due to the relaxation of single propanol molecules, supporting the commonly assumed cluster scenario for its explanation [10,24].

Coming back to $\Delta \ln \varepsilon''(\nu)$, the increase below the minimum frequency observed in Fig. 3(a) most likely points to nonlinear contributions from ionic charge transport. Small amounts of ionic impurities resulting in non-zero dc conductivity are nearly unavoidable. Via the relation $\varepsilon'' \propto \sigma'/\nu$, this leads to a minimum at the left flank of the loss peak and an $1/\nu$ divergence towards low frequencies. As seen in Fig. 3(a), the mentioned low-frequency increase of $\Delta \ln \varepsilon''(\nu)$ sets in at a frequency only slightly above that of the $\varepsilon''$ minimum revealed in Fig. 3(b). The nonlinear behavior of ionic conductivity is a well-known fact [42,43] and its detailed investigation is outside of the scope of the present work.

### 3.2. α-relaxation

The open arrows included in Fig. 3(a) indicate the α-relaxation rates $\nu_\alpha = 1/(2\pi\tau_\alpha)$ as determined from fits of low-field permittivity spectra [26]. The onsets of the above-mentioned shoulders, observed at frequencies beyond the peak of $\Delta \ln \varepsilon''(\nu)$, obviously arise at frequencies close to $\nu_\alpha$. This finding can be ascribed to the same mechanism as described in section 3.1 for the explanation of the high-field variation of $\varepsilon''$ in canonical dipolar glass formers [1,3]: At the high-frequency flank of the α peak, field energy is absorbed by the heterogeneous regions with relaxation rates $\nu \geq \nu_\alpha$, leading to an enhancement of $\varepsilon''$. This behavior could not be observed in the $\Delta \ln \varepsilon''$ spectra of 2-ethyl-1-butanol reported in [6] because, at the investigated temperature of 175 K, the α peak was outside of the covered frequency window [44]. However, in that work at least the onset of a decrease of $\Delta \ln \varepsilon''(\nu)$ at the highest investigated frequencies was already detected, which could not be explained by the employed model. In Fig. 3(a) a corresponding decrease for 1-propanol is observed, e.g., between about 0.1 - 2 Hz for 108 K. We ascribe this behavior to the mentioned strongly diminished ability to absorb field energy at the low-frequency flank of the α peak, known from simpler glass formers like glycerol [1,12]. In the frequency region where this reduction of $\Delta \ln \varepsilon''(\nu)$ is found, the α-relaxation peak obviously starts to dominate the detected loss (cf. Fig. 3(b)).

A signature of the nonlinear behavior of the α peak is also found in $\Delta \ln \varepsilon'(\nu)$: In the inset of Fig. 4, close to the α-relaxation rate a small peak shows up, caused by the same field-absorption effects as discussed for the loss. The canonical glass formers glycerol and propylene carbonate show similar behavior [45]. As mentioned above, for 5-methyl-3-heptanol and 4-methyl-3-heptanol the field variation of $\varepsilon'$ was reported in Ref. [11]. However, due to limited experimental resolution at high frequencies, no conclusions on the behavior in the α- or β-relaxation regime can be drawn from these data.

### 3.3. β-relaxation

Finally, an examination of $\Delta \ln \varepsilon''(\nu)$ at the highest investigated frequencies in Fig. 3(a) reveals a continuous decrease towards low values or even zero, at least for the two lowest presented temperatures, where a clear signature of the β relaxation shows up in Fig. 3(b). Interestingly, in glycerol and propylene carbonate similar behavior was found for the region of the so-called excess wing [12]. In the loss spectra, this spectral feature shows up as a second more shallow power law at the high-frequency flank of the α peak. It was shown [46], that the excess wing is caused by a secondary relaxation peak that is partly submerged under the dominating α peak. The β relaxation in 1-propanol can be assumed to be of Johari-Goldstein type [20,21,26,31]. The same was suggested for the secondary relaxation causing the excess wing in canonical dipolar glass formers like glycerol [46,47] (but also other opinions exist; see, e.g., [48]). Therefore the strong reduction or even absence of a nonlinear effect in the β-relaxation regime of 1-propanol can be assumed to have the same origin as the absence of nonlinearity in the excess-wing region of glycerol and propylene carbonate, reported in [12]. As discussed there, this finding would be, e.g., consistent with a recent theory relating nonlinear properties and molecular cooperativity [18]. Secondary relaxations are often assumed to be of non-cooperative nature, leading, e.g., to the Arrhenius temperature-dependence of their relaxation time, in contrast to the common deviation of $\tau_\alpha(T)$ from such temperature dependence, ascribed to cooperativity [13,49].

However, when considering the present results in the *β*-relaxation regime, it should be noted that a rather large number of high-field cycles may be needed to ensure that indeed the equilibrium dielectric response is measured [33]. In Ref. [50] it was shown that this effect may be especially critical in the excess-wing region. As noted in the Supplementary Information of Ref. [12], in the experiments on glycerol and propylene carbonate the actual attainment of equilibrium was ensured by a comparison of results with different cycle numbers. In the present measurements, for $T \geq 108$ K in the *β*-peak region (Fig. 3(a)) the time, during which the field was applied, was at least a factor of three longer than $\tau_\alpha$ (cf. section 2). By all means, this should lead to an equilibrium state [50]. However, for 100 K, where $\tau_\alpha$ is rather long ($\tau_\alpha \approx 30$ s [20,21,26]), this was not the case. If assuming a similarly slow approach of equilibrium in the *β*-relaxation regime as demonstrated for the excess wing in Ref. [50], these results therefore would not reflect steady-state properties. It is clear that more experimental work is necessary to clarify the nonlinear properties of the *β* relaxation, which currently are in progress in our group.

## 4. Summary and Conclusions

In summary, a detailed characterization of the field-induced variation of the dielectric constant and loss of 1-propanol was performed. Using microspheres as capacitor-spacer material, high fields of 468 kV/cm could be achieved, enabling the resolution of even relatively small nonlinear effects. Together with the rather broad covered frequency range, this allowed for the detection of the nonlinear behavior in the regimes of the Debye, *α*, and *β* relaxations. The behavior in the Debye regime is governed by two different mechanisms: At low frequencies, dielectric saturation leads to a pronounced reduction of $\varepsilon'$ and $\varepsilon''$. Obviously, in contrast to other monohydroxy alcohols [10,11], ring-chain conversions of the hydrogen-bonded molecule clusters that generate the Debye relaxation play no important role in 1-propanol. The magnitude of the observed saturation effect in 1-propanol is stronger than expected for a relaxation of single molecules, clearly pointing to the cluster-like nature of the relaxing entities. Our results seem to indicate smaller cluster sizes than in other monohydroxy alcohols, in agreement with the trends found in simulation studies [38,41]. In the frequency region of the right flank of the Debye peak, the absorption of field energy leads to an increase of both quantities, qualitatively similar to the behavior at the *α* relaxation of canonical glass formers [1,3,12].

Both the field-induced variations of $\varepsilon'$ and of $\varepsilon''$ show clear signatures of the *α* relaxation in 1-propanol. Overall, the present results nicely demonstrate that the *α* peak of this monohydroxy alcohol exhibits similar nonlinear properties as conventional glass formers, consistent with heterogeneous relaxation behavior [1,3]. Finally, in the region of the Johari-Goldstein *β*-relaxation of 1-propanol the observed nonlinear effects seem to strongly diminish or even vanish completely. This behavior agrees with the findings for the excess-wing regime of glycerol and propylene carbonate [12], supporting ideas that excess wing and Johari-Goldstein *β*-relaxation have the same origin.

Overall, the performed nonlinear dielectric measurements have revealed valuable information on the three types of relaxational processes observed in 1-propanol. Especially, it seems that such measurements are a promising tool to elucidate the nature of molecular clustering (e.g., chains, rings, branched aggregates). This may be used to distinguish the different classes of monohydroxy alcohols characterized by different molecular aggregates.

## Acknowledgements

This work was partly supported by the Deutsche Forschungsgemeinschaft via research unit FOR 1394.